\documentclass[preprintnumbers,amsmath,amssymb,aps,prl]{revtex4}
\usepackage{color}
\newcommand*{\eps}{{\rlap{\lower2ex\hbox{$\,\,\tilde{}$}}{\epsilon_{ijk}}}}
\newcommand*{\EPS}{{\rlap{\lower2ex\hbox{$\,\,\tilde{}$}}{\epsilon_{i'j'k'}}}}
\newcommand*{\lmq}{{\rlap{\lower2ex\hbox{$\,\,\tilde{}$}}{\epsilon_{lmq}}}}
\newcommand*{\jmq}{{\rlap{\lower2ex\hbox{$\,\,\tilde{}$}}{\epsilon_{jmq}}}}
\newcommand*{\jql}{{\rlap{\lower2ex\hbox{$\,\,\tilde{}$}}{\epsilon_{jql}}}}
\newcommand*{\jlm}{{\rlap{\lower2ex\hbox{$\,\,\tilde{}$}}{\epsilon_{jlm}}}}
\newcommand*{\imq}{{\rlap{\lower2ex\hbox{$\,\,\tilde{}$}}{\epsilon_{imq}}}}
\newcommand*{\iql}{{\rlap{\lower2ex\hbox{$\,\,\tilde{}$}}{\epsilon_{iql}}}}
\newcommand*{\ilm}{{\rlap{\lower2ex\hbox{$\,\,\tilde{}$}}{\epsilon_{ilm}}}}
\newcommand*{\lmn}{{\rlap{\lower2ex\hbox{$\,\,\tilde{}$}}{\epsilon_{lmn}}}}
\newcommand*{\abc}{{\rlap{\lower2ex\hbox{$\,\,\tilde{}$}}{\epsilon_{abc}}}}
\newcommand*{\N}{{\rlap{\lower2ex\hbox{$\,\,\tilde{}$}}{N}}}
\newcommand{\tN}{{\rlap{\lower2ex\hbox{$\,\,\tilde{}$}}{N}}}
\newcommand*{\tM}{{\rlap{\lower2ex\hbox{$\,\,\tilde{}$}}{M}}}
\newcommand*{\imn}{{\rlap{\lower2ex\hbox{$\,\,\tilde{}$}}{\epsilon_{imn}}}}
\newcommand*{\qt}{\ln q^{\frac{1}{3}}}

\begin{document}
\title{Intrinsic Time Quantum Geometrodynamics}

\author{Eyo Eyo Ita III}\email{ita@usna.edu}
\address{Physics Department, US Naval Academy. Annapolis, Maryland}
\author{Chopin Soo}\email{cpsoo@mail.ncku.edu.tw}
\address{Department of Physics, National Cheng Kung University, Taiwan}
\author{Hoi-Lai Yu}\email{hlyu@phys.sinica.edu.tw}
\address{Institute of Physics, Academia Sinica, Taiwan}

\bigskip

\begin{abstract}
Quantum Geometrodynamics with intrinsic time development and momentric variables is presented. An underlying $SU(3)$ group structure at each spatial point regulates the theory. The intrinsic time behavior of the theory is analyzed, together with its ground state and primordial quantum fluctuations. Cotton-York potential dominates at early times when the universe was small; the ground state naturally resolves Penrose's Weyl Curvature Hypothesis, and thermodynamic and gravitational `arrows of time'  point in the same direction. Ricci scalar potential corresponding to Einstein's General Relativity emerges as a zero-point energy contribution.
A new set of fundamental commutation relations without Planck's constant emerges from the unification of Gravitation and Quantum Mechanics.
\\
\\
\end{abstract}

\maketitle

\subsection{Intrinsic time development, and momentric as $SU(3)$ generators}

A century after the birth of Einstein's General Relativity (GR), successful quantization of the gravitational field remains the preeminent challenge.
Geometrodynamics bequeathed with positive-definite spatial metric is the simplest consistent framework to implement fundamental canonical commutation relations (CR) predicated on the existence of spacelike hypersurfaces.
In quantum gravity,  spacetime is a `concept of limited validity'\cite{Wheeler} and  ```time" must be determined intrinsically'\cite{DeWitt}. A full theory of Quantum Geometrodynamics dictated by first-order Schr\"{o}dinger evolution
 in intrinsic time, $i\hbar\frac{\delta\Psi}{\delta{T}}={H}_{\rm Phys.}\Psi$,  and
equipped with diffeomorphism-invariant physical Hamiltonian and time-ordering was formulated recently\cite{SOOYU, SOOYU1}.
Decomposition of the fundamental geometrodynamic degrees of freedom, $(q_{ij},\widetilde{\pi}^{ij})$,  singles out the canonical pair $(\ln q^{\frac{1}{3}}$, $\widetilde{\pi})$ which commutes with
the remaining unimodular $\overline{q}_{ij}=q^{-\frac{1}{3}}q_{ij}$, and traceless $\overline{\pi}^{ij}=q^{\frac{1}{3}}\bigl(\widetilde{\pi}^{ij}-\frac{1}{3}q^{ij}\widetilde{\pi}\bigr)$.
Hodge decomposition for compact manifolds yields
$\delta \qt = \delta{T}+\nabla_i\delta{Y}^i$, wherein the spatially-independent $\delta T$ is a 3-dimensional diffeomorphism invariant (3dDI) quantity which serves as the intrinsic time interval, whereas $\nabla_i\delta{Y}^i$ can be gauged away since
${\mathcal L}_{\delta{{\overrightarrow N}}}\qt = \frac{2}{3}\nabla_i{\delta N^i}$. The Hamiltonian, ${H}_{\rm Phys}=\int\frac{{\bar H}(x)}{\beta}d^3x$, and ordering of the time development operator
$U(T,T_0)={\bf T}\{{\exp}[-\frac{i}{\hbar}\int^T_{T_0}H_{\rm Phys}(T')\delta{T}']\}$ are 3dDI provided
\begin{eqnarray}
\label{INTRINSIC1}
{\bar H} = {\sqrt{ {\bar \pi^{ij}}  {\bar G_{ijkl}} {\bar \pi^{kl}} + \mathcal{V}[q_{ij}]}}
\end{eqnarray}\noindent
is a scalar density of weight one\cite{SOOYU}.
Einstein's GR (with $\beta= \frac{1}{\sqrt{6}}$ and ${\mathcal V} =- \frac{q}{(2\kappa)^2}[R - 2\Lambda_{\it{eff}} $]) is a particular realization of this wider class of theories.

Difficulties in implementing ${\bar\pi}^{ij}$ as self-adjoint traceless operator in the metric representation lead us to summon the momentric variable which is classically $\bar \pi^{i}_{j} = \bar q_{jm}\bar \pi^{im}$.
The fundamental CR is restriction of Klauder's affine algebra\cite{Klauder} to traceless momentric and unimodular part of the spatial metric,
\begin{equation}
\label{RELATIONS}
[ \bar q_{ij}(x), \bar q_{kl}(y)]=0,~~
[\bar q_{ij}(x), \hat{\bar{\pi}}^{k}_{l}(y)]= i\hbar\bar{E}^k_{l(ij)}\delta(x-y),~~
[ \hat{\bar{\pi}}^{i}_{j}(x), {\bar{\pi}}^{k}_{l}(y)]= \frac{i\hbar}{2}\bigl(\delta^k_j\hat{\bar{\pi}}^i_l-\delta^i_l\hat{\bar{\pi}}^k_j\bigr)\delta(x-y);
\end{equation}\noindent
wherein $\bar{E}^i_{j(mn)}=\frac{1}{2}\bigl(\delta^i_m\overline{q}_{jn}+\delta^i_n\overline{q}_{jm}\bigr)-\frac{1}{3}\delta^i_j\overline{q}_{mn}$
(with properties $\delta^{j}_{i}\bar{E}^i_{j(mn)} = \bar{E}^i_{j(mn)} \bar q^{mn}=0;\bar{E}^i_{jil}=\bar{E}^i_{jli}=\frac{5}{3}\overline{q}_{jl}$) is the vielbein for the
supermetric ${\bar G}_{ijkl} =  \bar{E}^m_{n(ij)}\bar{E}^n_{m(kl)}$.
Quantum mechanically, the momentric operators and CR can be explicitly realized  in the metric representation by
\begin{equation}
\hat{\bar{ \pi}}^{i}_{j}(x)=\frac{\hbar}{i}\bar{E}^i_{j(mn)}(x)\frac{\delta}{\delta \bar q_{mn}(x)} =\frac{\hbar}{i}\frac{\delta}{\delta \bar q_{mn}(x)}\bar{E}^i_{j(mn)}(x)=\hat{\bar{ \pi}}^{\dagger i}_{j}(x)
\end{equation}\noindent
which are self-adjoint on account of $[\frac{\delta}{\delta\bar{q}_{mn}(x)},\bar{E}^i_{j(mn)}(x)]=0$.
These eight momentric variables generate $SL(3,R)$ transformations  of $\bar{q}_{ij}$ which preserve positivity and unimodularity.
Moreover, it is crucial to realize they generate, by themselves, at each spatial point, an $SU(3)$ algebra. In fact, with Gell-Mann matrices $\lambda^{A=1,...,8}$ ,
$T^{A}(x)=  \frac{1}{\hbar\delta(0)}(\lambda^{A})^{j}_{i}\hat{\bar \pi}^{i}_{j}(x)$ satisfy
 $[T^{A}(x),T^{B}(y)]= i{f}^{AB}\,_CT^{C}\frac{\delta (x-y)}{\delta(0)}$ with $SU(3)$ structure constants $f^{AB}\,_C$\cite{Gell-Mann}.

Perturbative renormalizability of GR can be attained by adding higher derivative terms, but 4-covariance locks higher temporal and spatial derivatives to the
same order, compromising the stability and unitarity of the theory\cite{Horava}. Paradigm shift from 4-covariance to 3dDI not only resolves the `problem of time', but also leads
to the generic weight two (semi)positive-definite potential\cite{SOOYU},
\begin{equation}
{ \mathcal V}=[\frac{1}{2}( {\bar q_{ik}}{\bar q_{jl}} + {\bar q_{il}}{\bar q_{jk}} ) +\gamma {\tilde q_{ij}} {\tilde q_{kl}}] {\tilde W^{ij} }{\tilde W^{kl}},  \quad \gamma \geq -\frac{1}{3}; \qquad
{\tilde W^{i}_{j} }=[\sqrt q (\Lambda' + a' R)\delta^{i}_{j} + b\hbar\sqrt q {\bar R}^{i}_{j} +g\hbar\tilde C^{i}_{j} ];
\end{equation}\noindent
wherein $R$  and ${\bar R}^i_j$ are respectively the scalar and traceless parts of the spatial Ricci curvature, while ${\tilde C}^i_j$ is the Cotton-York tensor (density) which is third order in spatial derivatives and associated with dimensionless coupling constant $g$.
In conjunction with intrinsic time evolution with $H_{\rm Phys}$,  this framework presents,  in quantum gravity, a new vista to surmount conceptual and technical obstacles.

\subsection{Free Hamiltonian}
The free theory  is characterized by $SU(3)$ invariance generated by the momentric (whereas ${\tilde\pi}^{ij}$ generate translations which do not preserve the positivity of the metric), because the Casimir invariant $T^AT^A$  is related to the kinetic operator in Eq.(\ref{INTRINSIC1})  through
\begin{equation}
\frac{\hbar^{2}\delta^{2}(0)}{2} T^{A}T^{A}={\hat{\bar \pi}}^{i\dagger }_{j} {\hat{\bar \pi}}^{j}_{i}= {\hat {\bar \pi}}^{i}_{j}{ \hat {\bar \pi}}^{j}_{i} ={\hat {\bar \pi}^{ij}}  {\hat{\bar G}_{ijkl}} {\hat {\bar \pi}}^{kl} .
\end{equation}
\noindent
The upshot is its spectrum can be labeled by eigenvalues of the complete commuting set at each spatial point comprising the two
Casimirs $L^{2}=T^{A}T^{A}, C=d_{ABC}T^AT^BT^C \propto \det(\hat{\bar \pi}^{i}_{j})$,  Cartan subalgebra $T^{3},T^{8}$, and isospin $I=\sum_{B=1}^3 T^{B}T^{B}$.   An underlying group structure has the advantage the action of momentric on wavefunctions  by functional differentiation can be traded for its well defined action as generators of $SU(3)$ on states expanded in this basis since
\begin{equation}
\label{CASIMIR}
\frac{\hbar}{i}(\lambda^{A})^{i}_{j}\bar{E}^j_{i(mn)}\frac{\delta}{\delta \bar{q}_{mn}(x)}
\langle\bar{q}_{kl}|\prod_{y}{| l^2, C, I, m_{3}, m_{8}\rangle}_y
=\frac{\hbar\delta(0)}{2}\langle\bar{q}_{kl}|T^{A}(x)\prod_{y}{| l^2, C, I, m_{3}, m_{8}\rangle}_y .
\end{equation}
For the free theory, the ground state with zero energy, $|0\rangle$,  corresponds to $l^2=0 \,\forall x$, which is an $SU(3)$  singlet state annihilated by all the momentric operators (${\hat{\bar \pi}}^{i}_{j}(x)|0\rangle =0$); and it is  also 3dDI because $-2\nabla_j{\hat{\bar \pi}}^{j}_{i}$ generates spatial diffeomorphisms of ${\bar q}_{ij}$.

\subsection{Asymptotic behavior of the Hamiltonian at early and late intrinsic times}

Hodge decomposition for $\delta \qt$ and its Heisenberg equation of motion lead to $\frac{d}{dT} \qt(x, T)= \frac{\partial}{\partial T}\qt  + \frac{1}{i\hbar}[\qt, H_{\rm Phys}] = 1$; with solution
$\ln[\frac{q(x, T) }{q(x,T_{\rm now})}]=3{(T-T_{\rm now})},$ $~ -\infty < T < \infty$. Moreover, the Hodge decomposition also implies the change in the global intrinsic time is proportional to the logarithmic change in the volume of the universe, $\delta T = \frac{2}{3}\delta \ln V$ i.e. $T-T_{\rm now} = \frac{2}{3}\ln(V/V_{\rm now})$\cite{redshift}. Explicitly separating out T-dependence from entities (labeled with overline) which depend  only on ${\bar q}_{ij}$,
\begin{eqnarray}
\label{INTRINSIC2}
{\tilde W^{i}_{j} }&=&[\sqrt q (\Lambda' + a' R)\delta^{i}_{j} + b' \sqrt q {\bar R}^{i}_{j} +g\hbar\tilde C^{i}_{j} ]\nonumber\\
&=&[\sqrt q (\Lambda' + a' q^{-\frac{1}{3}}{\bar q}^{kl}{\overline R}_{kl})\delta^{i}_{j} + b' {\sqrt q}q^{-\frac{1}{3}}\bar q^{ik}\overline{\bar R}_{kj} +g\hbar\tilde C^{i}_{j} ] + (\partial_{i}\ln q \,\,{\rm terms}),
\end{eqnarray}
with $q$-independent Cotton York tensor density $\tilde C^{i}_{j}$ which is conformally invariant.
The theory is not (intrinsic)time-reversal invariant; furthermore, exponential scaling behavior of $q$ with intrinsic time implies  in the limit $ T -T_{\rm now} \rightarrow -\infty,  V/V_{\rm now} \rightarrow 0 $ (i.e. early times when the universe was very small in volume), the potential $\mathcal{V}$  was dominated by the Cotton-York term, whereas the limit $ T -T_{\rm now} \rightarrow \infty,  V/V_{\rm now} \rightarrow \infty $ (i.e. late times when the universe becomes large) will be dominated by the cosmological constant term. This is compatible with current observations of our ever expanding universe, with a middle period in which curvature and cosmological terms are comparable in importance.
\indent

\subsection{Early universe and Cotton-York dominance}

In the era of Cotton-York dominance at the beginning of the universe, $\bar H =
\sqrt{\hat{\bar{\pi}}^{\dagger j}_i \hat{\bar{\pi}}^{i}_j+g^2\hbar^2\tilde{C}^j_i\tilde{C}^i_j}$.
A number of intriguing facts conspire to simplify and regulate the Hamiltonian:
The  traceless  Cotton-York tensor density is expressible as ${\tilde C}^i_j =\bar{E}^i_{j(mn)}\frac{\delta W}{\delta \bar q_{mn}} $,
wherein $W=\frac{1}{4}\int{\tilde\epsilon}^{ijk}({\bar\Gamma}^l_{im} \partial_j{\bar\Gamma}^m_{kl} +\frac{2}{3}{\bar\Gamma}^l_{im}{\bar\Gamma}^m_{jn}{\bar\Gamma}^n_{kl})\,d^3x$ is the 3dDI Chern-Simon functional of the affine connection of ${\bar q}_{ij}$. This leads to the similarity transformation of the momentric,
\begin{equation}
{\hat Q}^{i}_{j}= e^{gW}{\hat{\bar \pi}}^{i}_{j}e^{-gW}=\frac{\hbar}{i}{\bar E}^i_{j(mn)}[\frac{\delta}{\delta{\bar q}_{mn}}- g\frac{\delta W}{\delta {\bar q}_{mn}}]=\frac{\hbar}{i}{\bar E}^i_{j(mn)}\frac{\delta}{\delta{\bar q}_{mn}}+ i g\hbar{\tilde C}^i_j .
\end{equation}
\noindent
Moreover, $[\hat{\bar{\pi}}^i_j,\tilde{C}^j_i]=0$ i.e. without  zero point energy (ZPE) contribution\cite{footnote2}. Consequently, the Hamiltonian density is simply
$\bar H = {\sqrt {{\hat Q}^{\dagger i}_{j}{\hat Q}^{j}_{i}}}=  {\sqrt {{\hat Q}^{i}_{j}{\hat Q}^{\dagger j}_{i}}}$. While ${\hat Q}^{\dagger i}_{j}$ and ${\hat Q}^{i}_{j}$ are related to $\hat{\bar{\pi}}^i_j$ by $e^{\mp gW}$  similarity transformations, they are non-Hermitian and generate two unitarily inequivalent representations of the non-compact group $SL(3,R)$ at each spatial point; whereas the momentric  $\hat{\bar{\pi}}^i_j = \frac{1}{2}({\hat Q}^{\dagger i}_{j} +{\hat Q}^{i}_{j})$  generates a unitary representation of ${\prod}_{x }SU(3)_x$.

\subsection{Initial state of the universe and Penrose's Weyl Curvature Hypothesis}

From the classical perspective, $\bar H =
\sqrt{{\bar{\pi}}^{j}_i {\bar{\pi}}^{i}_j+g^2\hbar^2\tilde{C}^j_i\tilde{C}^i_j}$ attains its lowest value iff the momentric and Cotton-York tensor vanish identically, the latter being precisely the criterion for conformal flatness in three dimensions\cite{footnote1}.
The vanishing of the momentric (hence traceless part of the classical extrinsic curvature) and spatial conformal flatness at $T \rightarrow -\infty (q \rightarrow 0)$ realize a Robertson-Walker Big Bang compatible with Penrose's hypothesis that the initial singularity must have vanishing 4-dimensional Weyl curvature tensor. A solar mass black hole has a Bekenstein-Hawking entropy of $\sim 10^{21}$ per baryon.  By Penrose's estimate, with $10^{80}$ baryons in our universe, thermalization of gravitational degrees of freedom at the initial hot Big Bang would imply an entropy of ${{10}^{123}}$. `Our extraordinarily special Big Bang' with low entropy\cite{Penrose} emerges naturally from the ground state of $H_{Phys}$ in the era of Cotton-York dominance; and 2nd Law thermodynamic `arrow of time' and `gravitational arrow of intrinsic time' (of increasing volume) point in the same direction.

Vanishing of both the traceless momentric and Cotton-York tensor implies the trace of the momentum $\pi \propto {\bar H}$ also vanishes, the extrinsic curvature is thus totally absent, which is the junction condition needed for Euclidean continuation of the metric (for instance, continuation to Euclidean $S^4$  at the conformally flat $S^3$ section at the throat of Lorentzian de Sitter metric).  The quantum context may be in  agreement with  Hartle-Hawking `no-boundary proposal' for the wavefunction of the universe\cite{Hartle-Hawking}. But it should be noted the intrinsic time framework discussed here already allows a continuation of $\beta$ in $H_{\rm Phys}$ to imaginary values and Euclidean partition functions; moreover, from the formula of the emergent lapse function\cite{SOOYU}, imaginary $\beta$ leads to emergent semi-classical space-times which are Euclidean in signature.

The Cotton-York interaction is introduced through the extension $\hat{\bar{\pi}}^i_j \rightarrow e^{gW}{\hat{\bar \pi}}^{i}_{j}e^{-gW} ={\hat Q}^{i}_{j}$; thus ${\hat Q}^{i}_{j}$ and ${\hat Q}^{\dagger i}_{j}$ respectively annihilate
the state $e^{\pm gW}|0\rangle$. Moreover both these states are annihilated by $\bar H $ because ${ {{\hat Q}^{\dagger i}_{j}{\hat Q}^{j}_{i}}}=  {{{\hat Q}^{i}_{j}{\hat Q}^{\dagger j}_{i}}}$, due to the absence of Cotton-York ZPE.  So any linear combination $Ae^{gW}|0\rangle + Be^{-gW}|0\rangle = N\cosh[g(W -W_o)]|0\rangle$ is a zero energy state.
From the quantum perspective, the classical conformally flat configuration with vanishing Cotton-York tensor is the extremum of $W$, and thus precisely a saddle point for the ground state wavefunction $\Psi_0[{\bar q}_{ij}]= N\cosh [g(W  -W_o)]\langle {\bar q}_{ij}|0\rangle$  (with the simplest choice of constant $\langle {\bar q}_{ij}|0\rangle$).
Expanding $W$ about the saddle point $\frac{\delta W}{\delta{\bar q}_{ij}} ={\tilde C}^{ij} =0$ leads to
\begin{equation}
W[{\bar q}_{ij}] -W_o =  W[\Gamma ({\tilde C}^i_j =0)]  + \frac{1}{2}\int d^3x\int d^3y\, \delta{\bar q}_{ij}(x) H^{ijkl}(x,y)|_{{\tilde C}^m_n=0}\delta{\bar q}_{kl}(y)  + ... - W_o
\end{equation}
wherein $H^{ijkl}(x,y) =\frac{\delta^2 W}{\delta{\bar q}_{ij}(x) \delta{\bar q}_{kl}(y)}$ is the Hessian; and it is natural to choose $W_o$ to cancel the zeroth order term $ W[\Gamma ({\tilde C}^i_j =0)] $
 which is the Chern-Simons functional of a conformally flat connection\cite{Soo1}. $W_o= W[\Gamma ({\tilde C}^i_j =0)]$  is a topological entity, invariant under infinitesimal variations of the metric.  The ground state will thus contain primordial quantum fluctuations which can be studied. For instance, the Hessian, which for flat metric is $-\frac{1}{2}\delta^{jk}\epsilon^{iml}\partial_m\partial^2\delta(x-y)$, is the inverse of the two-point correlation function, and Cotton-York dominance would thus be compatible with $ \sim 1/k^3$  behavior.
 With $|N|^{-2}  = \int [D{\bar q} ]\cosh ^2[g(W -W_o)] $ the wavefunction is normalized, but its definition involves the computation of  $\int [D{\bar q} ] e^{\pm 2g[(W -W_o)]}$ which are just partition functions of Chern-Simons actions\cite{footnote4}.

\subsection{Emergence of Einstein-Hilbert Gravity }

Ricci curvature terms become increasingly important in the potential after the initial era of Cotton-York dominance.
They can be introduced in a manner which preserves the underlying structure which regulate the Hamiltonian by extending the Chern-Simons action with
3dDI invariants of the spatial metric. This not only guarantees 3dDI invariance; but also makes the Hamiltonian density the square-root of a (semi)positive-definite and self-adjoint object
${\hat Q}^{\dagger i}_{j}{\hat Q}^{j}_{i}$; and ensures the preservation of all these properties even under renormalization of the coupling constants. In increasing order of spatial derivatives, these invariants are
$ \Lambda\int {\sqrt q} d^3x , EH =b\int {\sqrt q}R d^3x$, and the Chern-Simons functional of the affine connection with dimensionless coupling constant. Even higher derivative curvature invariants will come along with super-renormalizable dimensional coupling constants, while the cosmological constant volume term commutes with ${\hat{\bar \pi}}^{i}_{j}$ due to the traceless projector $\bar{E}^i_{j(mn)}$. To wit, only the Einstein-Hilbert action in 3 dimensions and the Chern-Simons functional
 are relevant i.e. total $W_T=\frac{g}{4}\int{\tilde\epsilon}^{ijk}({\bar\Gamma}^l_{im} \partial_j{\bar\Gamma}^m_{kl} +\frac{2}{3}{\bar\Gamma}^l_{im}{\bar\Gamma}^m_{jn}{\bar\Gamma}^n_{kl})\,d^3x + b\int {\sqrt q}Rd^3x$.
 This leads to
\begin{equation}
{\hat Q}^{i}_{j}= e^{W_T}{\hat{\bar \pi}}^{i}_{j}e^{-W_T}=\frac{\hbar}{i}{\bar E}^i_{j(mn)}[\frac{\delta}{\delta{\bar q}_{mn}} - \frac{\delta W_T}{\delta{\bar q}_{mn}}]=\frac{\hbar}{i}{\bar E}^i_{j(mn)}\frac{\delta}{\delta{\bar q}_{mn}} +ib\hbar\sqrt{q}{\bar R}^i_j   + ig\hbar{\tilde C}^i_j
\end{equation}
wherein (again due to the ${\bar E}^i_{j(mn)}$ projector) only the traceless part of the Ricci tensor remains. The Hamiltonian density is then
\begin{eqnarray}
\bar H = {\sqrt {{\hat Q}^{\dagger i}_{j}{\hat Q}^{j}_{i}}} = \sqrt{\hat{\bar{\pi}}^{\dagger j}_i \hat{\bar{\pi}}^{i}_j+\hbar^2(g\tilde{C}^i_j + b\sqrt{q}{\bar R}^i_j)(g\tilde{C}^j_i +b\sqrt{q}{\bar R}^j_i ) +[\hat{\bar{\pi}}^{i}_j, ib\hbar\sqrt{q}{\bar R}^j_i]}\,;
\end{eqnarray}
wherein the ZPE from incorporating the  Einstein-Hilbert action in $W_T$ is  $[\hat{\bar{\pi}}^{i}_j, ib\sqrt{q}\hbar{\bar R}^j_i] =-\frac{5}{12}b\hbar^2\delta(0){\sqrt q}(5{R}-\frac{9}{\epsilon})$\cite{Heatkernel}.
Remarkably, the potential for Einstein's theory, which is the Ricci scalar, and a (positive) c-number term emerge. This means the simple Hamiltonian density  ${\sqrt {{\hat Q}^{\dagger i}_{j}{\hat Q}^{j}_{i}}}$ (with all its aforementioned advantages) already contains Einstein's GR with cosmological constant. Furthermore, ${\bar R}^i_j$ and the Cotton-York tensor only appear in the higher-curvature higher-derivative combination $(g\tilde{C}^j_i +a\sqrt{q}{\bar R}^j_i )(g\tilde{C}^i_j+a\sqrt{q}{\bar R}^i_j) $ (these `non-GR' terms are automatically absent in homogeneous FLRW cosmology (that the Weyl Curvature Hypothesis holds in the Cotton-York era has been addressed), and also in constant curvature slicings of Painlev\'{e}-Gullstrand solutions of black holes\cite{Lin}). Consequently,  except for Cotton-York preponderance at very early times, Einstein's GR dominates at low curvatures and long wavelengths
in a theory in which `four-dimensional symmetry is not a fundamental property of the physical world'\cite{Dirac}.

\subsection{Quantum Geometrodynamics redux}

  Local $SL(3,R)$ transformations of ${\bar q}_{kl}$ are generated through $U^\dagger(\alpha) {\bar q}_{kl}(x) U(\alpha) = (e^{\frac{\alpha(x)}{2}})^m_k {\bar q}_{mn}(x) (e^{\frac{\alpha(x)}{2}})^n_l$,  wherein $ U(\alpha)=e^{-\frac{i}{\hbar}\int \alpha^i_j {\bar \pi}^j_i d^3y}$\cite{Klauder}; while the generator of spatial diffeomorphisms for the momentric and unimodular spatial d.o.f. is effectively $ D_i =-2{\nabla}_j{\bar\pi}^j_i$,
with smearing $\int N^i D_i d^3x =\int (2{\nabla}_jN^i){\bar\pi}^j_i  d^3x $ after integration by parts\cite{footnote3}. The action of spatial diffeomorphisms can thus be subsumed by specialization to $\alpha^i_j =2\nabla_jN^i$, with the upshot
that $SL(3,R)$ transformations which are {\it not} spatial diffeomorphisms are parametrized by $\alpha^i_j$ complement to $2{\nabla}_jN^i$. Given a background metric ${q}^B_{ij}= q^{\frac{1}{3}}{\bar q}^B_{ij}$, this complement is
 precisely characterized by the choice of  transverse traceless (TT) parameter  $(\alpha_{TT})^i_j  := {q}^B_{jk}\alpha^{(ik)}_{Phys}$  (because the condition $\nabla_{B}^j(\alpha_{TT})^i_j=0 $ excludes non-trivial $N^i$  through $\nabla^2_{B}N^i=0$ if $(\alpha_{TT})^i_j $ were of the form $2\nabla^B_j N^i$ (the label $B$ denotes the connection of $q^B_{ij}$)). TT conditions impose 4 restrictions on the symmetric $\alpha^{(ij)}_{Phys}(x)$, leaving exactly 2 free parameters. The action of $U_{Phys}( \alpha_{TT}) = e^{-\frac{i}{\hbar}\int (\alpha_{TT})^i_j {\bar \pi}^j_i d^3x}$  (which is thus local $SL(3,R)$ modulo spatial diffeomorphism) on any 3dDI wavefunction would result in an inequivalent state. The caveat is TT conditions require a particular background metric to be defined. However, in Ref.\cite{SOOITA} a basis of infinitely squeezed states was explicitly realized by Gaussian wavefunctionals $\Psi[\bar q]_{{q}^B} \propto \exp[-\frac{1}{2}\int {\tilde f}_{\epsilon}({\bar q}_{ij}-{\bar q}^B_{ij}) {\bar G}^{ijkl}_B ({\bar q}_{kl}-{\bar q}^B_{kl}) d^3x]$.  3dDI is recovered in the limit of zero Gaussian width with divergent $\hbox{lim}_{\epsilon\rightarrow{0}}{\tilde f}_{\epsilon}\rightarrow \delta(0)$. These  localized Newton-Wigner states are infinitely peaked at $q^B_{ij}$ which can be deployed to actualize the TT conditions.  The action of $U_{Phys}( \alpha_{TT})$ on these states would thus generate 2 infinitesimal local physical excitations at each spatial point.

In the preceding discussions, the entity $\delta(0)$ which denotes the 3-dimensional coincidence limit, $\lim_{x\rightarrow y}\delta(x-y)$, was left untouched, with the understanding that it can be regularized, for instance, by normalized Gaussians of infinitesimal but non-zero width. However, the underlying $SU(3)$ structure already provides unambiguous guidance on how to regularize the theory. The Hamiltonian assumes the elegant form,
 \begin{equation} H_{Phys}={\hbar}\int \sqrt{\Big(e^{-W_T} {T^A(x)}e^{W_T}\Big)\Big(e^{W_T}{T^A(x)}e^{-W_T}\Big)}\,\frac{\delta(0)}{{\sqrt 2}\beta}d^3x;
  \end{equation}
  wherein  $\frac{\delta(0)}{\beta}d^3x$ is a dimensionless volume element, its divergence to be absorbed by renormalization of $\beta$\cite{footnoteD}.
  With the cancelation of $\hbar$ on both sides of the Schr\"odinger equation, our universe is described by a fundamental equation with dimensionless Hamiltonian and intrinsic time. What is paramount to causality is not the actual dimension of time (an exemplar is intrinsic time interval measured with dimensionless redshift in FLRW cosmology), but the sequence and ordering in time. Even as $\hbar$  will continue to leave its imprints in physics in the conversion factor between $SU(3)$ generators $T^A$ and the momentric (hence momentum of the gravitational field), unification of gravitation and quantum mechanics comes with the demotion of its elementary significance.
  With dimensionless fundamental variables, the CR are\cite{efootnote}
  \begin{equation}
\label{RELATIONS}
[ \bar q_{ij}(x), \bar q_{kl}(y)]=0,~~
[\bar q_{ij}(x), T^A(y)]= \frac{i}{2}\Big((\lambda^A)^k_i{\bar q}_{kj} + (\lambda^A)^k_j{\bar q}_{ki}\Big)\frac{\delta (x-y)}{\delta(0)};~~
[T^{A}(x),T^{B}(y)]= i{f}^{AB}\,_CT^{C}\frac{\delta (x-y)}{\delta(0)}.
\end{equation}\noindent
Quantum essence is still embodied in the non-commutativity, but Planck's constant is absent.

\section*{Acknowledgements}

This work was supported in part by the US Naval Academy, Annapolis, Maryland; the Ministry of Science and Technology (R.O.C.) under Grant No.
NSC101-2112-M-006 -007-MY3; and the Institute of Physics, Academia Sinica (R.O.C).  H.-L. Yu would like to thank the Yukawa Institute
for Theoretical Physics for partial support and hospitality during the early stage of this work.


\begin{thebibliography}{99}
\bibitem{redshift}In FLRW cosmology, intrinsic time interval can be measured through the redshift $z$.
\bibitem{footnote2}The commutator is a scalar density of weight 2, with the density weight carried by $\lim_{x \rightarrow y}\delta (x-y)$ and the Levi-Civita tensor density from
${\tilde C}^{ij} \propto {\tilde \epsilon}^{(imn}\nabla_mR^{j)}_n$. Contraction over spatial indices of  ${\tilde \epsilon}^{imn}$ with ${q}_{mn}$ and $R_{mn}$, and $\epsilon^{ijk}\nabla_i\nabla_j\nabla_k$ all vanish, the latter due to Bianchi identity. Explicit computations verify the vanishing of the commutator.
\bibitem{footnote1}Cotton-York dominance at very early times is robust against the inclusion  in $\mathcal V$ of usual Yang-Mills fields or fermionic matter, because the corresponding Hamiltonian densities contain positive powers of $q (T) \sim e^{3(T-T_{\rm now})}$. Conventional scalar field potentials also contain positive powers of $q$, but the weight two kinetic term ${\tilde\pi}_\phi^2$ is independent of $q$. However, this does not change the criterion of vanishing momentric and Cotton-York tensor for the classical lowest energy state which then requires ${\tilde\pi}_\phi^2 =0$ as well.
\bibitem{Heatkernel}
Explicit computations yield $[\hat{\bar{\pi}}^{i}_j, ib\sqrt{q}\hbar{\bar R}^j_i] = -{5b\hbar^2}{\sqrt q}(\frac{\delta(0)}{3} R  + \frac{1}{2}\nabla^2\delta(0))$.
The heat kernel, $K(\epsilon;x,y)$ with $ \hbox{lim}_{\epsilon\rightarrow{0}}K(\epsilon;x,y)=\delta(x-y)$, presents, in the coincidence limit and with infinitesimal $\epsilon$,  the means to regularize $\nabla^2\delta(0)$ for generic metrics.
In terms of Seeley-DeWitt coefficients $b_n$, $2\sigma(x,y)$ the square of the geodesic length, $\Delta_V$ the Van Vleck determinant,
$K(\epsilon;x,y)=(4\pi\epsilon)^{-3/2}\Delta^{1/2}_V(x,y)\sqrt{q(y)}e^{-\sigma(x,y)/2\epsilon}\sum_{n=0}^{\infty}b_n(x,y;\nabla^2)\epsilon^n$, wherein $\epsilon$ is of dimension $L^2$. Thus the heat kernel equation implies $\nabla^2K(\epsilon;x,y)=\frac{\partial K(\epsilon;x,y)}{\partial\epsilon}\rightarrow -(\frac{3}{2\epsilon}-b_1)\delta(0)$ in the $x=y$ and infinitesimal $\epsilon$ limit. Seeley-DeWitt coefficient $b_1 = \frac{R}{6}$ yields
$[\hat{\bar{\pi}}^{i}_j, ib\sqrt{q}\hbar{\bar R}^j_i] =-\frac{5}{12}b\hbar^2\delta(0){\sqrt q}(5{R}-\frac{9}{\epsilon})$.
\bibitem{footnoteD}It is conventional for the spatial label $x$ to carry the dimension of length, even though it is non-dynamical and a mere dummy variable to be integrated over. But it makes more physical sense to take partial derivatives, and calculate curvatures, with respect to a dimensionless variable $X= x/L$. To wit, $W_T =  gW[\Gamma(X)] + (bL)\int \sqrt{ q(X)}R(X) d^3X $, with dimensionless coupling constant $(bL)$; furthermore, $\frac{\delta(0)}{\beta}d^3x =\frac{1}{\beta}[\lim_{y\rightarrow x}\delta(y-x)]d^3x = \frac{1}{\beta}[\lim_{Y \rightarrow X}\delta(Y-X)]d^3X$.
\bibitem{efootnote}With ${\bar q}_{ij} = {\bar e}_{ai }{\bar e}^{a}_j$,  the fundamental CR for unimodular dreibein is  $[{\bar e}_{ai}(x), T^A(y)] =i(\frac{\lambda^A}{2})^k_i{\bar e}_{ak}\frac{\delta (x-y)}{\delta(0)}$, with 8 independent components in both ${\bar e}_{ai}$ and $T^A$.
\end{thebibliography}
\end{document}